\documentclass{elsart}

\usepackage{cite}
\usepackage{amssymb,amsmath,bm}
\usepackage{psfrag,graphicx}

\newcommand\DS\displaystyle
\newlength{\mpwidth}
\newlength{\sfigwidth}
\newlength{\figwidth}
\begin{document}

\begin{frontmatter}
\title{Baptista-type chaotic cryptosystems: Problems and countermeasures}
\thanks{Citation information: \textit{Physics Letters A}, 332(5-6):368-375, 2004.}
\author[hk1]{Shujun Li\corauthref{corr}},
\author[hk1]{Guanrong Chen},
\author[hk2]{Kwok-Wo Wong},
\author[xa]{Xuanqin Mou} and
\author[xa]{Yuanlong Cai}
\address[hk1]{Department of Electronic Engineering, City University of Hong Kong,
83 Tat Chee Avenue, Kowloon Tong, Hong Kong, China}
\address[hk2]{Department of Computer Engineering and Information Technology, City University of Hong Kong,
83 Tat Chee Avenue, Kowloon Tong, Hong Kong, China}
\address[xa]{School of Electronics and Information Engineering,
Xi'an Jiaotong University, Xi'an, Shaanxi 710049, China}
\corauth[corr]{The corresponding author, e-mail address:
\texttt{hooklee@mail.com}, web site:
\texttt{http://www.hooklee.com}.}

\begin{abstract}
In 1998, M.~S. Baptista proposed a chaotic cryptosystem, which has
attracted much attention from the chaotic cryptography community:
some of its modifications and also attacks have been reported in
recent years. In [Phys. Lett. A 307 (2003) 22], we suggested a
method to enhance the security of Baptista-type cryptosystem,
which can successfully resist all proposed attacks. However, the
enhanced Baptista-type cryptosystem has a nontrivial defect, which
produces errors in the decrypted data with a generally small but
nonzero probability, and the consequent error propagation exists.
In this Letter, we analyze this defect and discuss how to rectify
it. In addition, we point out some newly-found problems existing
in all Baptista-type cryptosystems and consequently propose
corresponding countermeasures. \iffalse It will be shown that
Baptista-type chaotic cryptosystem can be effectively enhanced to
achieve an acceptable overall performance in practice.\fi
\end{abstract}
\begin{keyword}
chaos \sep encryption \sep cryptanalysis \sep Baptista-type
chaotic cryptosystem

\PACS 05.45.Ac/Vx/Pq
\end{keyword}

\end{frontmatter}

\section{Introduction}

In \cite{Baptista:ChaoticCipher:PLA98}, M.~S. Baptista proposed a
chaotic cryptosystem based on partitioning the visiting interval
of chaotic orbits of the logistic map. After its publication,
several modified versions have been proposed
\cite{Wong:ChaoticCipher:CPC2001, Wong:ChaoticCipher:PLA2002,
Wong:ChaoticCipher&Hashing:PLA2003, Wong:ShortCiphertext:PLA2003,
Palacios:CyclingChaosCryptography:PLA2002,
LiShujun:ChaoticCipher:PLA2003}. On the other hand, some attacks
have been reported as tools of breaking the original Baptista-type
cryptosystem and some of its modified versions
\cite{Jakimoski:ChaoticCryptanalysis:PLA2001,
Alvarez:BreakingBapistaCipher:PLA2003,
Alvarez:BreakingWongCipher:CPC2003,
Alvarez:BreakingWongCipher:PLA2004}. In this section, we give a
brief survey on Baptista-type chaotic cryptosystems, including the
original scheme and some modified versions, and on some proposed
attacks. In the following sections, we will show some problems of
this class of cryptosystems and then propose some countermeasures
for enhancing its overall performance.

%\subsection{Baptista-type chaotic cryptosystems: the original and modified versions}

At first, we give a detailed introduction to the original
Baptista-type cryptosystem, as a basis of the whole Letter. Note
that different notations from those in
\cite{Baptista:ChaoticCipher:PLA98} are used to make the
description simpler and clearer.

Given a one-dimensional chaotic map $F:X\to X$ and an interval
$X'=[x_{\mathrm{min}},x_{\mathrm{max}})\subseteq X$, divide $X'$
into $S$ $\epsilon$-intervals: $\forall i=1\sim S$,
$X_i'=[x_{\mathrm{min}}+(i-1)\varepsilon,x_{\mathrm{min}}+i\varepsilon)$,
where $\varepsilon=\dfrac{x_{\mathrm{max}}-x_{\mathrm{min}}}{S}$.
Assume that plain messages are composed by $S$ different
characters, $\alpha_1,\cdots,\alpha_S$, and use a bijective map,
\begin{equation}
f_S:\bm{X_\epsilon}=\{X_1',\cdots,X_i',\cdots,X_S'\}\to
\bm{A}=\{\alpha_1,\cdots,\alpha_i,\cdots,\alpha_S\},
\end{equation}
to associate the $S$ different $\epsilon$-intervals with the $S$
different characters. By introducing an extra character
$\beta\not\in\bm{A}$, we can define a new function $f_S':X\to
\bm{A}\cup\{\beta\}$ as follows:
\begin{equation}
f_S'(x)=
\begin{cases}
f_S(X_i'), & x\in X_i',\\
\beta, & x\notin X'.
\end{cases}
\end{equation}

Based on the above notations, for a plain-message
$M=\{m_1,m_2,\cdots,m_i,\cdots\}$ ($m_i\in \bm{A}$), the original
Baptista-type cryptosystem can be described as follows.

\begin{itemize}
\item \textit{The employed chaotic system}: the logistic map,
$F(x)=bx(1-x)$.

\item \textit{The secret key}: the association map $f_S$, the
initial condition $x_0$ and the control parameter $b$ of the
logistic map.

\item \textit{The encryption procedure}: a) initialize
$x_0^{(0)}=x_0$; b) encrypt the $i$-th plain-character $m_i$ as
follows: iterate the chaotic system from $x_0^{(i-1)}$ to find a
chaotic state $x$ satisfying $f_S'(x)=m_i$, record the iteration
number $C_i$ as the $i^{th}$ cipher-message unit and
$x_0^{(i)}=F^{C_i}\left(x_0^{(i-1)}\right)=F^{C_1+C_2+\cdots+C_i}(x_0)$.

\item \textit{The decryption procedure}: for each cipher-message
unit $C_i$, iterate the chaotic system for $C_i$ times from
$x_0^{(i-1)}$, and then use
$x_0^{(i)}=F^{C_i}\left(x_0^{(i-1)}\right)$ to derive the current
plain-character as follows: $m_i=f_S'\left(x_0^{(i)}\right)$.

\item \textit{Constraints on $C_i$}: each cipher-message unit
$C_i$ should satisfy $N_0\leq C_i\leq N_{\mathrm{max}}$ ($N_0=250$
and $N_{\mathrm{max}}=65532$ in
\cite{Baptista:ChaoticCipher:PLA98}). Since there exist many
options for each $C_i$ in $[N_0,N_{\mathrm{max}}]$, an extra
coefficient $\eta\in[0,1]$ is used to choose the right number: if
$\eta=0$, $C_i$ is chosen as the minimal number satisfying
$f_S'(x)=m_i$; if $\eta\neq 0$, $C_i$ is chosen as the minimal
number satisfying $f_S'(x)=m_i$ and $\kappa\geq \eta$
simultaneously, where $\kappa$ is a pseudo-random number with a
normal distribution within the interval $[0,1]$.
\end{itemize}

The original Baptista-type chaotic cryptosystem has the following
four defects.
\begin{enumerate}
\item The distribution of the ciphertext is non-uniform, and the
occurrence probability decays exponentially as $C_i$ increases
from $N_0$ to $N_{max}$ (see Fig. 3 of
\cite{Baptista:ChaoticCipher:PLA98} and also Fig. 1 of
\cite{Wong:ChaoticCipher:CPC2001}).

\item At least $N_0$ chaotic iterations are needed to encrypt a
plain-character, which makes the encryption speed very slow as
compared with most conventional ciphers.

\item The ciphertext size is larger than the plaintext size.

\item It is insecure against some different attacks proposed in
\cite{Jakimoski:ChaoticCryptanalysis:PLA2001,
Alvarez:BreakingBapistaCipher:PLA2003}, since some useful
information about the chaotic system can be obtained from the
ciphertext $\{C_i\}$, i.e., the iteration numbers of the chaotic
system.
\end{enumerate}

In recent years, some modifications have been proposed as possible
remedies for the above defects \cite{Wong:ChaoticCipher:CPC2001,
Wong:ChaoticCipher:PLA2002, Wong:ChaoticCipher&Hashing:PLA2003,
Wong:ShortCiphertext:PLA2003,
Palacios:CyclingChaosCryptography:PLA2002,
LiShujun:ChaoticCipher:PLA2003}. Meanwhile, cryptanalysis works
have also been developed to break some modifications
\cite{Alvarez:BreakingWongCipher:CPC2003,
Wong:Reply4Breaking:CPC2003, Alvarez:BreakingWongCipher:PLA2004}.

In \cite{Wong:ChaoticCipher:CPC2001}, the first modified version
was proposed to overcome the first defect of the original
Baptista-type cryptosystem. According to
\cite{Alvarez:BreakingWongCipher:CPC2003,
Wong:Reply4Breaking:CPC2003}, this modified version is still
insecure against the keystream attack proposed in
\cite{Alvarez:BreakingBapistaCipher:PLA2003}.

In \cite{Wong:ChaoticCipher:PLA2002,
Wong:ChaoticCipher&Hashing:PLA2003}, to overcome the second
defect, the original Baptista-type cryptosystem was enhanced by
dynamically updating the association map $f_S$. However, following
the cryptanalysis given in
\cite{Alvarez:BreakingWongCipher:PLA2004}, the two modified
versions are still insecure, since the essential security defect
(i.e., the existence of $C_i$ in the ciphertext) remains. In
\cite{Wong:ShortCiphertext:PLA2003}, utilizing the technique
proposed in \cite{Wong:ChaoticCipher:PLA2002,
Wong:ChaoticCipher&Hashing:PLA2003}, another modified version was
further proposed to achieve shorter ciphertext. This modification
has not been cryptanalyzed, but the attacks proposed in
\cite{Alvarez:BreakingWongCipher:PLA2004} may be generalized to
break it.

In \cite{Palacios:CyclingChaosCryptography:PLA2002}, as a new idea
of increasing the security, cycling chaos generated by multiple
different chaotic attractors is used instead of chaos generated
from one single chaotic map. Though the use of multiple chaotic
maps can effectively increase the complexity of some attacks, it
seems that the keystream attack proposed in
\cite{Alvarez:BreakingBapistaCipher:PLA2003} may still work to its
advantage.

In \cite{LiShujun:ChaoticCipher:PLA2003}, we proposed a new
modification to essentially enhance the security of the original
Baptista-type cryptosystem. In this scheme, the original
ciphertext stream $\{C_i\}$ is masked by a pseudo-random number
stream and then be output as the final ciphertext stream. In this
case, it is impossible for an attacker to get the number of
chaotic iterations from the ciphertext, so that all proposed
attacks will fail. Unfortunately, later we noticed that this
modified scheme has a nontrivial defect, which produces errors in
the decrypted data with a generally small but nonzero probability.
In the next section, we give more details on this defect and
discuss how to rectify it.

In all the above Baptista-type cryptosystems, there exist some
general problems that have not been reported before, which can
influence the overall performance of the cryptosystems to some
extent. In Sec. \ref{section:NewProblems} of this Letter, we will
further discuss these problems and provide some corresponding
countermeasures.

\section{Rectifying our early-proposed remedy of Baptista-type chaotic
cryptosystem that can resist all proposed attacks}

\subsection{A brief introduction of the enhanced Baptista-type cryptosystem}

Since the occurrence of $C_i$ in the ciphertext stream is the
prerequisite of all proposed attacks, we can bypass it by
concealing $C_i$ in the ciphertext stream. A natural idea is to
secretly mask $C_i$ with a pseudo-random number stream. It is easy
to generate the pseudo-random number stream from the chaotic
system itself. Given a pseudo-random number generation function
$f_{be}(\cdot)$, using $\oplus$ to denote the masking operation,
the enhanced Baptista-type cryptosystem proposed in
\cite{LiShujun:ChaoticCipher:PLA2003} can be described as follows
(without changing other details of the original cryptosystem, such
as the constraints on $C_i$):

\begin{itemize}
\item \textit{The encryption procedure}: for the $i$-th
plain-character $m_i$, iterate the chaotic system starting from
$x_0^{(i-1)}$ to find a suitable chaotic state $x$ satisfying
$f_S'(x)=m_i$, record the number of chaotic iterations starting
from $x_0^{(i-1)}$ to $x$ as $\widetilde{C}_i$ and
$x_0^{(i)}=x=F^{\widetilde{C}_i}\left(x_0^{(i-1)}\right)$. Then,
the $i$-th cipher-message unit of $m_i$ is
$C_i=\widetilde{C}_i\oplus f_{be}\left(x_0^{(i)}\right)$.

\item \textit{The decryption procedure}: for each ciphertext unit
$C_i$, firstly iterate the chaotic system for $N_0$ times and set
$\widetilde{C}_i=N_0$, then perform the following operations: if
$\widetilde{C}_i\oplus f_{be}(x)=C_i$ then use the current chaotic
state $x$ to derive the plain-character $m_i$ and goto the next
ciphertext unit $C_{i+1}$; otherwise, iterate the chaotic system
once and $\widetilde{C}_i++$, until the above condition is
satisfied.

\item \textit{The selection of $f_{be}(\cdot)$}: due to the
non-uniformity of the ciphertext, it has been known that
$f_{be}(\cdot)$ cannot be freely selected to avoid information
leaking. For example, the simplest function $f_{be}(x)=x$ is not
secure. Two classes of such functions are suggested, and both can
make information leaking impossible. If the distribution of $C_i$
is modified to be uniform with some techniques\footnote{As
mentioned in \cite{LiShujun:ChaoticCipher:PLA2003}, two methods
are available: the modification proposed in
\cite{Wong:ChaoticCipher:CPC2001} and the entropy-based lossless
compression technique \cite{Castleman:DigitalImageProcessing96}.},
then $f_{be}(\cdot)$ can freely selected.
\end{itemize}

\subsection{A defect in the above modified Baptista-type cryptosystem}
\label{section:OurDefect}

Although the above modified Baptista-type cryptosystem can resist
the attacks proposed in
\cite{Jakimoski:ChaoticCryptanalysis:PLA2001,
Alvarez:BreakingBapistaCipher:PLA2003}, considering
$\widetilde{C}_i\oplus f_{be}(x)=\widetilde{C}_i'\oplus
f_{be}(x')$ is possible for $\widetilde{C}_i\neq
\widetilde{C}_i'$, erroneous plain-characters may be ``decrypted"
with a generally small but nonzero probability: at the decipher
side, when $\widetilde{C}_i\oplus f_{be}(x)=C_i$, the restored
``$\widetilde{C}_i$" may not be the real $\widetilde{C}_i$ at the
encipher side, so that the restored chaotic state $x$ is wrong
and, as a result, the decrypted plain-character is also wrong.

At first, let us see how serious this defect is. We can estimate
the error probability at the encipher side as follows. Apparently,
the decryption is correct if and only if the real
$\widetilde{C}_i$ never occur before the first $x$ satisfying
$f'_S(x)=m'$ is found. That is, for a specific $\widetilde{C}_i$,
the probability to successfully restore $\widetilde{C}_i$ (i.e.
the probability to get the correct decryption) via the above
decryption procedure is
\begin{eqnarray}
P_c\left(\widetilde{C}_i\right) & = &
P\left\{\bigcap\nolimits_{k=N_0}^{\widetilde{C}_i-1}\left(k\oplus
f_{be}\left(F^k\left(x_0^{(i-1)}\right)\right)\neq C_i\right)\right\}\nonumber\\
& = &
P\left\{\bigcap\nolimits_{k=N_0}^{\widetilde{C}_i-1}\left(f_{be}\left(F^k\left(x_0^{(i-1)}\right)\right)\neq
k\oplus C_i\right)\right\}.
\end{eqnarray}
Generally, assume the bit size of $C_i$ is $n$ (for the original
Baptista-type cryptosystem $n=16$) and the chaotic orbit
$\left\{F^k\left(x_0^{(i-1)}\right)\right\}$ has a uniform
distribution, we have: $\forall
C_i,P\left\{f_{be}\left(F^k\left(x_0^{(i-1)}\right)\right)=C_i\right\}=2^{-n}$,
i.e.,
\begin{equation}
P\left\{f_{be}\left(F^k\left(x_0^{(i-1)}\right)\right)\neq k\oplus
C_i\right\}=1-2^{-n}.
\end{equation}
Assume $f_{be}\left(F^k\left(x_0^{(i-1)}\right)\right)=k\oplus
C_i(k=N_0\sim \widetilde{C}_i-1)$ are independent events. Then, we
can deduce
$P_c\left(\widetilde{C}_i\right)=\left(1-2^{-n}\right)^{\widetilde{C}_i-N_0}$.
It is obvious that $P_c\left(\widetilde{C}_i\right)\to 0$ as
$\widetilde{C}_i\to\infty$, which means any decryption behaves
like a random guess after a sufficiently long period of time.

\begin{figure}
\centering
\includegraphics[width=0.5\textwidth]{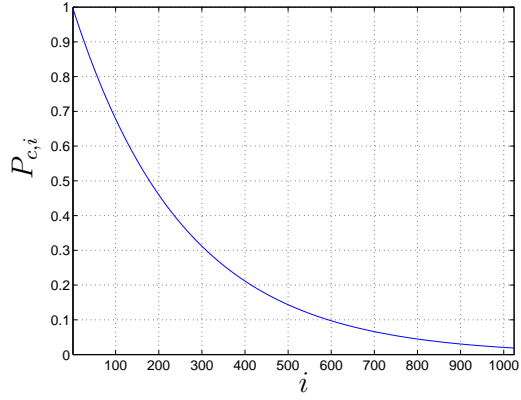}
\caption{$P_{c,i}$ with respect to the position of the
plain-character $i$.} \label{figure:Pci}
\end{figure}

Considering the non-uniform distribution of $\widetilde{C}_i$, for
the first plain-character $m_1$, from the total probability rule
we can calculate\footnote{Here, assume $P\{C_i>N_{max}\}=0$ (see
Sec. \ref{section:BaptistaCipher} for an explanation).} the final
probability $P_{c,1}$:
\begin{eqnarray}
P_{c,1} & = &
\sum_{k=N_0}^{N_{max}}P\left\{\widetilde{C}_i=k\right\}\cdot
P_c(k)\nonumber\\
& = &
\sum_{k=N_0}^{N_{max}}P\left\{\widetilde{C}_i=k\right\}\cdot\left(1-2^{-n}\right)^{k-N_0}.
\end{eqnarray}
To simplify the calculation, without loss of generality, assume
$F(x)$ visits each $\epsilon$-interval with the same
probability\footnote{logistic map does not satisfy this
requirement, so we suggest using PWLCM to replace the logistic map
in Sec. \ref{section:LogisticMapProblems}.} $p=1/S$. Then, we have
$P\{\widetilde{C}_i=k\}=p(1-p)^{k-N_0}$, so that
\begin{eqnarray}
P_{c,1} & = &
\sum_{k=N_0}^{N_{max}}p(1-p)^{k-N_0}\cdot(1-2^{-n})^{k-N_0}\nonumber\\
& = & \sum_{k'=0}^{N_{max}-N_0}p\cdot
q^{k'}=p\cdot\frac{1-q^{N_{max}-N_0}}{1-q},
\end{eqnarray}
where $q=(1-p)\cdot(1-2^{-n})$. When
$S=256,n=16,N_0=250,N_{max}=65532$ (values in the original
Baptista-type cryptosystem), $P_{c,1}\approx 0.9961240899211138$.
Considering $1/(1-P_{c,1})\approx 258$, we expect that one
plaintext with wrong leading plain-character will occur averagely
in 258 plain-characters. Here, note that all plain-characters
after a wrong plain-character will be wrong with a high
probability close to 1, i.e., there exists error propagation. It
is obvious that the error propagation makes things worse for
$i>1$:
\begin{equation}
P_{c,i}=\left(\prod\limits_{j=1}^{i-1}P_{c,j}\right)\cdot\dfrac{p\left(1-q^{N_{max}-N_0}\right)}{1-q}=\left(\prod\limits_{j=1}^{i-1}P_{c,j}\right)\cdot
P_{c,1}=P_{c,1}^i.
\end{equation}
For the above calculated $P_{c,1}$, $P_{c,i}$ with respect to $i$
is shown in Fig.~\ref{figure:Pci}. As $i$ increases, the
probability decreases exponentially. Once $P_{c,i}$ goes below
$1/S$, a random guess process will replace the role of the
designed decipher.

\subsection{Rectification to the existing defect}

Now, we try to rectify the above-discussed encryption/decryption
scheme to avoid the existing defect. The goal is to ensure that
$\forall i$, $P_{c,i}\equiv 1$.

With a memory unit allocated to store $N_{max}-N_0+1$ variables
$B[N_0]\sim B[N_{max}]$ representing $C_i=N_0\sim C_i=N_{max}$
respectively, we propose to change the encryption/decryption
procedure as follows:

\begin{itemize}
\item \textit{The encryption procedure}: for the $i$-th
plain-character $m_i$, firstly set $B[N_0]=\cdots=B[N_{max}]=0$,
iterate the chaotic system starting from $x_0^{(i-1)}$ for $N_0$
times, set $\widetilde{C}_i=N_0$, and then perform the following
operations: $C_i=\widetilde{C}_i\oplus f_{be}(x)$, $B[C_i]++$, if
the current chaotic state $x$ satisfying $f_S(x)=m_i$, then a
2-tuple ciphertext $(C_i,B[C_i])$ is generated and set
$x_0^{(i)}=x$ and then goto the next plain-character $m_{i+1}$;
otherwise, repeat this procedure until a ciphertext is generated.

\item \textit{The decryption procedure}: for each ciphertext unit
$(C_i,B_i)$, firstly iterate the chaotic system for $N_0$ times
and set $\widetilde{C}_i=N_0$, then perform the following
operations: if $\widetilde{C}_i\oplus f_{be}(x)=C_i$ for the
$B_j$-th times then use the current chaotic state $x$ to derive
the plain-character $m_i$ and goto the next ciphertext unit
$(C_{i+1},B_{i+1})$; otherwise iterate the chaotic system and
$\widetilde{C}_i++$ for 1 iteration, until the above condition is
satisfied.
\end{itemize}

\begin{figure}[!htbp]
\centering
\begin{minipage}[b]{\mpwidth}
\centering
\includegraphics[width=\sfigwidth]{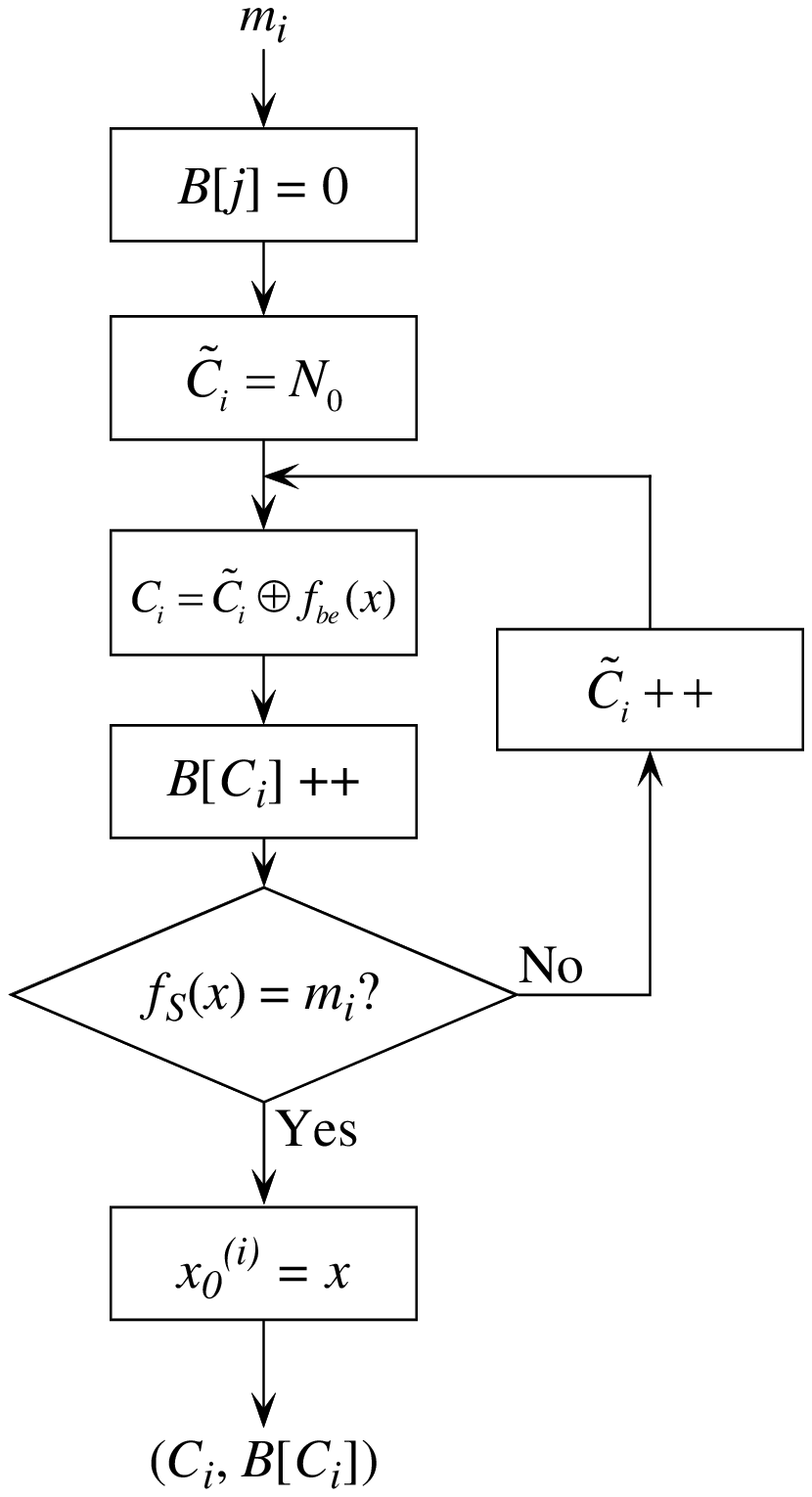}\\
a) Encryption procedure
\end{minipage}
\begin{minipage}[b]{\mpwidth}
\centering
\includegraphics[width=\sfigwidth]{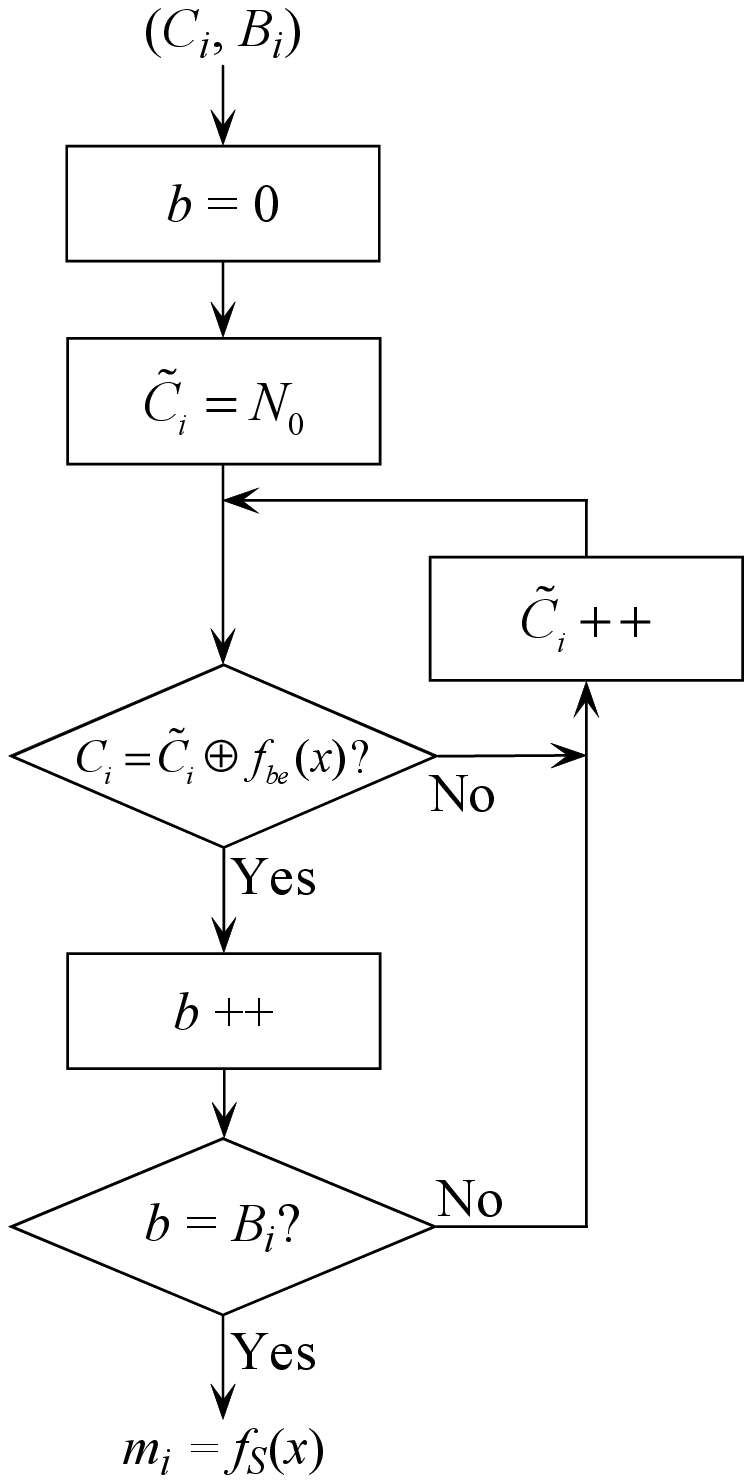}\\
b) Decryption procedure
\end{minipage}
\caption{The encryption and decryption procedures of the rectified
Baptista-type cryptosystem.} \label{figure:Baptista:Enhance}
\end{figure}

In Fig.~\ref{figure:Baptista:Enhance}, we show flow charts for the
above rectified encryption and decryption procedures, in which
$B[j]=0$ means setting all $B[j]\;(j=N_0\sim N_{max})$ to zeros,
$\widetilde{C}_i'=N_0$ denotes $N_0$ chaotic iterations and
setting $\widetilde{C}_i'$ to $N_0$, and $\widetilde{C}_i'++$
indicates one chaotic iteration and increasing $\widetilde{C}_i'$
by one.

Compared with the original Baptista-type cryptosystem, this
rectified cryptosystem manages to solve the aforementioned defect
with a cost of adding more implementation complexity:
\begin{enumerate}
\item Extra memory is needed to store $N_{max}-N_0+1$ variables
$B[j]$. When each $B[j]$ is stored as a 2-byte integer, the memory
size is $2\times(N_{max}-N_0+1)$ bytes. When $N_{max}=65532$ and
$N_0=250$, it is not greater than 128 KB.

\item The encryption speed becomes lower since $N_{max}-N_0+1$
variables $B[j]$ should be set to zero for each plain-character.

\item The ciphertext size becomes even longer: $B[C_i]$ is added
into each ciphertext unit.
\end{enumerate}

Fortunately, the requirement on extra memory is acceptable in all
digital computers nowadays (128 KB is not so much for a computer
with over tens or hundreds of MB in memory), and the encryption
speed will not be influenced much when this rectified cipher is
implemented in hardware with parallel support: all $N_{max}-N_0+1$
variables $B[j]$ can be set to zeros within a clock cycle
simultaneously, which eliminates the negative effect on the
encryption speed. In addition, chaotic iteration can be run in
parallel with $C_i=\widetilde{C}_i\oplus f_{be}(x)$, $B[C_i]++$
and $f_S(x)=m_i?$ with pre-calculation and delay design.
Therefore, the above rectification is quite practical in enhancing
the performance of Baptista-type cryptosystem. Moreover, the
enlargement of the ciphertext size can be effectively minimized by
some other methods, which will be discussed in the next
subsection.

\subsection{Minimizing the enlargement of the ciphertext size}
\label{section:MakingCiphertextShorter}

In the rectified cryptosystem, the ciphertext size is prolonged.
Some methods can be used to overcome this problem. Here, we
introduce two of them.

The first method is to use variable-length ciphertext. For
example, we can change the ciphertext as follows:
\begin{itemize}
\item When $B[C_i]=1$ and $N_0\leq C_i<N_{max}$, output $C_i$ as
the ciphertext.

\item When $B[C_i]=1$ and $C_i=N_{max}$, output $(N_{max},0)$ as
the ciphertext.

\item When $B[C_i]>1$, output $(N_{max},B[C_i],C_i)$ as the
ciphertext.
\end{itemize}
Assume the size of $C_i$ is $n$. We can calculate the mathematical
expectation of the ciphertext size, corresponding to one
plain-character, as follows:
\begin{equation}
(1-P_{c,1})\cdot\left(P\left\{N_0\leq\widetilde{C}_i<N_{max}\right\}\cdot
n+P\left\{\widetilde{C}_i=N_{max}\right\}\cdot
2n\right)+P_{c,1}\cdot 3n.
\end{equation}
Since $P\left\{\widetilde{C}_i=N_{max}\right\}\ll
P\left\{N_0\leq\widetilde{C}_i<N_{max}\right\}$, it can be
approximately reduced to
\begin{equation}
(1-P_{c,1})\cdot n+P_{c,1}\cdot 3n=(1+2P_{c,2})\cdot n.
\end{equation}
Generally, $0\approx P_{c,1}\ll 1$, so it is only a little bit
greater than $n$, which is the ciphertext size of the original
Baptista-type cryptosystem.

Another method is to use the compression algorithm suggested in
\cite{LiShujun:ChaoticCipher:PLA2001,
LiShujun:ChaoticCipher:PLA2003}. Since both $C_i$ and $B[C_i]$
have exponentially decreasing distributions, it is natural to use
lossless entropy-based compression algorithms to make the
ciphertext size shorter. Following the deduction given in
\cite{LiShujun:ChaoticCipher:PLA2001}, assuming that the bit size
of $C_i$ is $n$, the average size of the compressed $C_i$ will be
$n/2$. Since generally $0\approx P_{c,1}\ll 1$, it is obvious that
the average size of a compressed $B[C_i]$ will be close to 1 from
a probabilistic point of view. That is, the average ciphertext
size corresponding to one plaintext will be close to $n+1$.

Actually, we can also combine the above two methods to obtain a
better solution. Using a compressed $C_i$ in the first method can
successfully reduce the average ciphertext size to about $n/2$.

\section{Some general problems of Baptista-type chaotic cryptosystems and some corresponding countermeasures}
\label{section:NewProblems}

\subsection{Problems of the logistic map for encryption}
\label{section:LogisticMapProblems}

In the original Baptista-type chaotic cryptosystem and all its
modifications proposed thus far, the logistic map is used as the
chaotic system. But the logistic map is not a good chaotic system
for encryption due to the following reasons.

\textit{a) Non-uniform visiting probability on each
$\epsilon$-interval}. It is well-known that the logistic map has a
non-uniform invariant density function, which cause the visiting
probability of each $\epsilon$-interval to be different.
Experimental data given in Fig.~2 of
\cite{Baptista:ChaoticCipher:PLA98} have shown such a
disadvantage, but Baptista \cite{Baptista:ChaoticCipher:PLA98} did
not consider it as a negative factor to security. From a
cryptographical point of view, this issue indeed is not desirable
and may be vulnerable to some subtle statistics-based attacks. In
fact, such a disadvantage has been successfully utilized to design
an entropy-based attack by Alvarez et al. in
\cite{Alvarez:BreakingBapistaCipher:PLA2003}.

\textit{b) Limits on the control parameter $b$}. It is also
well-known that the logistic map becomes chaotic when
$b>3.5699\cdots$ and is completely chaotic (with the Lyapunov
exponent being maximal) only when $b=4$. To ensure that the
generated orbit is sufficiently chaotic, $b$ has to be
sufficiently close to 4, which limits the key space to be a small
set near 4. In addition, dynamics of the logistic map with
different values of the control parameter $b$ are different, which
may be utilized to develop some new attacks. In
\cite{LiShujun:ChaoticCipher:PLA2001}, we have shown a similar
defect in the chaotic cryptosystem developed in
\cite{Alvarez:ChaoticCipher:PLA}.

To avoid the above problems of the logistic map, we suggest using
the following piecewise linear chaotic maps (PWLCM) with the
\textit{onto} property \cite[\S3.2.1]{ShujunLi:Dissertation2003}
to replace the logistic map. An \textit{onto} PWLCM is generally
chaotic and has the following good dynamical properties on its
defining interval $X$\cite{ShujunLi:Dissertation2003,
Li:DPWLCM:IJBC2004, Baranovsky:PLCM:IJBC95,
Lasota:StochasticChaos97}: 1) its Lyapunov exponent
$\lambda=-\sum_{i=1}^m\|C_i\|\cdot\ln\|C_i\|$ satisfying
$0<\lambda<\ln m$; 2) it is exact, mixing and ergodic; 3) it has a
uniform invariant density function,
$f(x)=1/\|X\|=1/(\beta-\alpha)$; 4) its auto-correlation function
$\tau(n)=\frac{\DS 1}{\DS
\sigma^2}\lim_{N\rightarrow\infty}\frac{\DS 1}{\DS
N}\sum_{i=0}^{N-1}(x_i-\bar{x})(x_{i+n}-\bar{x})$ approaches zero
as $n\rightarrow\infty$, where $\bar{x},\sigma$ are the mean value
and the variance of $x$, respectively. A typical example is the
well-known skew tent map with a single control parameter
$p\in(0,1)$:
\begin{equation}
F(x)=
\begin{cases}
x/p, & x\in[0,p],\\
(1-x)/(1-p), & x\in(p,1].
\end{cases}
\end{equation}

Besides the above properties, PWLCM are also the simplest chaotic
maps from the digital implementation point of view. In addition,
some theoretical results on a direct digital realization of such
maps has been rigorously established \cite{Li:DPWLCM:IJBC2004},
which are useful for optimizing the implementation of
Baptista-type chaotic cryptosystems.

\subsection{Problems of the secret key}

In the original Baptista-type cryptosystem, the association map
$f_S$ also serves as part of the whole secret key. But we believe
that $f_S$ should not be included in the secret key from an
implementation consideration: it is too long for most users to
remember. If a secret algorithm is used to generate $f_S$, then
the secret key will be changed from $f_S$ to the key of the secret
algorithm, which is easier to implement.

In \cite{Alvarez:BreakingBapistaCipher:PLA2003}, the correlation
between $b$ and $x_0$ has been used to develop some theoretical
attacks. To avoid potential dangers, it is advisable to use only
control parameter(s) as the secret key.

\subsection{Dynamical degradation of digital chaotic systems}

In all versions of Baptista-type chaotic cryptosystems, dynamical
degradation of digital chaotic systems is neglected. However, it
has been found that dynamics of chaotic systems can easily
collapse in the digital world, and the dynamical degradation may
make some negative influences on the performance of digital
chaos-based applications \cite{ShujunLi:Dissertation2003,
Li:DPWLCM:IJBC2004}. Also, dynamical degradation may enlarge
differences among different visiting probabilities of different
$\epsilon$-intervals of a chaotic map.

Therefore, some methods should be used to improve such dynamical
degradation of the employed chaotic system in all Baptista-type
chaotic cryptosystems, which will ensure the visiting probability
of each $\epsilon$-interval to be close enough to the theoretical
value. As we discussed in \cite{ShujunLi:Dissertation2003,
Li:DPWLCM:IJBC2004}, a pseudo-random perturbation algorithm is
desirable and hence is recommended: use a simple pseudo-random
number generator (PRNG) to generate a small signal, to perturb the
concerned chaotic orbit every $\Delta\geq 1$ iterations.

\subsection{A trivial problem when $C_i>N_{max}$}
\label{section:BaptistaCipher}

The original Baptista-type cryptosystem did not consider what one
should do if $C_i>N_{\mathrm{max}}$. It seems to presume that
$C_i$ will never be greater than $N_{\mathrm{max}}$. However, this
is obviously not true. Here, assume $F(x)$ visits each
$\epsilon$-interval with the same probability, $p=1/S$. We can
deduce that
\begin{equation}
P\{C_i>N_{\mathrm{max}}\}=P\{C_i-N_0>N_{\mathrm{max}}-N_0\}=(1-p)^{N_{\mathrm{max}}-N_0}.
\end{equation}
Although this probability is very small when $N_{\mathrm{max}}$ is
large enough, it is nevertheless non-zero. To make the
cryptosystem rigorously complete, we propose to use the following
$(n+1)$-tuple data to replace $C_i$ when $C_i\geq
N_{\mathrm{max}}$:
$(\overbrace{N_{\mathrm{max}},\cdots,N_{\mathrm{max}}}^n,c_i)$,
where the number of total chaotic iterations is equal to
$C_i=N_{\mathrm{max}}\times n+c_i$. Apparently,
$(\overbrace{N_{\mathrm{max}},\cdots,N_{\mathrm{max}}}^n,c_i)$ can
be represented in a more brief format: $(N_{\mathrm{max}},n,c_i)$.
When $C_i=N_{\mathrm{max}}$, the 3-tuple ciphertext
$(N_{\mathrm{max}},n,c_i)$ can be further reduced to
$(N_{\mathrm{max}},0)$.

In fact, it is also acceptable to modify the original cryptosystem
as follows: once $C_i=N_{\mathrm{max}}$ occurs, immediately output
a 2-tuple data $(N_{\mathrm{max}},m_i)$ instead of $C_i$.
Considering $P\{C_i>N_{\mathrm{max}}\}$ is very small, such a tiny
chance of information leaking does no harm on the security of the
cryptosystem in practice.

\begin{ack}
The authors would like to thank Prof. L. Kocarev for his e-mails
to the first author, which motivated the authors to carefully
review their previous work reported in
\cite{LiShujun:ChaoticCipher:PLA2003} thereby leading to the
discovery of the defect discussed in Sec. \ref{section:OurDefect}
of this Letter. This research was partially supported by the
Applied R\&D Centres of the City University of Hong Kong under
grants no. 9410011 and no. 9620004.
\end{ack}

%\nocite{*}
\bibliographystyle{elsart-num}
\bibliography{PLA13800}

\end{document}